\documentclass[iop,apj,twocolumn,twocolappendix,numberedappendix]{emulateapj}
\usepackage{natbib}
\usepackage{amsfonts}
\usepackage{amsmath}
\usepackage{graphicx,epsfig}
\usepackage{footnote}
\newcommand{\Halp}{H${\alpha}$}

\newcommand{\ergscm}{\,erg\,cm$^{-2}$\,s$^{-1}$}

\newcommand{\ergs}{\,erg\,s$^{-1}$}

\newcommand{\Msol}{M$_{\odot}$}

\newcommand{\Rstar}{R$_{\mathrm *}$}
\newcommand{\Dstar}{D$_{\mathrm *}$}
\newcommand{\Teff}{$T_{\rm eff}$\,}

\newcommand{\degree}{\degr}
\newcommand{\Lbol}{\rm L$_{\rm bol}$}

\newcommand{\Lx}{$\rm L_{\rm X}$}

\newcommand{\gcas}{$\gamma$-Cas}

\shorttitle{The origin of the puzzling hard X-ray emission of $\gamma$ Cassiopeiae}
\shortauthors{Motch, Lopes de Oliveira \& Smith}

\begin{document}


\title{The origin of the puzzling hard X-ray emission\\ of $\gamma$ Cassiopeiae}


\author{Christian Motch\altaffilmark{1}, Raimundo Lopes de Oliveira\altaffilmark{2,3} and Myron A. Smith\altaffilmark{4}}

\affil{\altaffilmark{1}Observatoire Astronomique, Universit\'e de Strasbourg, CNRS, UMR 7550, 11 rue de l'Universit\'e, F67000 Strasbourg, France; christian.motch@unistra.fr\\ \altaffilmark{2}Universidade Federal de Sergipe, Departamento de F\'isica, Av. Marechal Rondon s/n, 49100-000 S\~ao Crist\'ov\~ao, SE, Brazil\\ \altaffilmark{3}Observat\'orio Nacional, Rua Gal. Jos\'e Cristino, 20921-400, Rio de Janeiro, Brazil\\ \altaffilmark{4}National Optical Astronomy Observatory, 950 N. Cherry Ave., Tucson, AZ, USA}

\begin{abstract}

Massive B and Be stars produce X-rays from shocks in high velocity winds with temperatures of a few million degrees and maximum X-ray luminosities of $\approx$ 10$^{31}$\ergs. Surprisingly, a sub-group of early Be stars exhibits $\geq$ 20 times hotter X-ray temperatures and $\geq$ 10 times higher X-ray luminosities than normal. This group of Be stars, dubbed $\gamma$-Cas analogs, contains about 10 known objects. The origin of this bizarre behavior has been extensively debated in the past decades. Two mechanisms have been put forward, accretion of circumstellar disk matter onto an orbiting white dwarf, or magnetic field interaction between the star and the circumstellar disk \citep{smith1999}. We show here that the X-ray and optical emissions of the prototype of the class, \gcas, are very well correlated on year time scales with no significant time delay. Since the expected migration time from internal disk regions that emit most of the optical flux to the orbit of the companion star is of several years, the simultaneity of the high energy and optical fluxes variations indicates that X-ray emission arises from close to the star. The systematic lack of magnetic field detection reported in recent spectro-polarimetric surveys of Be stars is consistent with the absence of strong magnetic wind braking in these fast spinning stars but put strong constraints on the possible origin of the magnetic field. We propose that in \gcas\ the magnetic field emerges from equatorially condensed subsurface convecting layers  the thickness of which steeply increases with rotation rate and that \gcas\ and its analogs are the most massive and closest to critical rotation Be stars.      

\end{abstract}

%%%%%%%%%%%%%%



\section{Introduction}

\gcas, the central star of the well known W shaped northern sky Cassiopeia constellation is a massive and hot B0.5IVe star  (M\,=\,16\,$\pm$\,2\,\Msol; \Teff\,$\approx$\,27,000\,K; \cite{stee2012}) surrounded by a circumstellar disk whose matter escaped from the star. The disk radiates strong emission lines of hydrogen that together with the hot stellar continuum define the spectral class of Be stars. \gcas\ was the first identified Be star \citep{secchi1866}. 

Although the optical properties of \gcas\ do not display marked peculiarities, the star stands out from the crowd of other Be stars by displaying an X-ray luminosity of a few 10$^{32}$\ergs\ (0.3$-$10\,keV), more than one order of magnitude brighter than most B and Be stars. The thermal X-ray spectrum is dominated by an exceedingly high temperature component with $kT$\,$\approx$\,13\,keV \citep{lopes2010}, extending up to $\approx$\,100\,keV \citep{shrader2015}, compared to $kT$\,$\approx$\,0.5\,keV in normal B and Be stars \citep{cohen1997}. In addition, the X-ray light-curve of \gcas\ exhibits a unique complex pattern of variability involving slow modulations on time scales of a few hours down to shot noise like events as short as a few seconds (see \cite{white1982,lopes2010,smith2012} and references therein). \gcas\ experienced a close to normal B star episode in the early 1940's and since then has more of less steadily built a new decretion disk and increased its optical brightness \citep{doazan1983,henry2012}. Low amplitude quasi-periodic oscillations (QPOs) with periods in the range of  70\,$\pm$\,20\,d \citep{smith2006} dominate the optical variability of the star on shorter time scales. Galactic X-ray surveys have revealed a growing number (at least 10 so far) of $\gamma$-Cas analogs \citep{lopes2006,motch2007,nebot2013,rauw2013,smith2015}. 

The origin of this outstanding X-ray emission affecting a small fraction of Be stars remains mysterious. Two distinct hypotheses have been so far proposed. The discovery that \gcas\ is a 203.6\,d binary with a $\approx$\,1\Msol\ companion star in a nearly circular orbit  \citep{harmanec2000,miro2002,nemravova2012,smith2012} supported the idea arising from some similarities to classical Be/X-ray binaries, that accretion of matter onto a compact object, most probably a white dwarf, could power the unusual X-ray luminosity \citep{white1982,haberl1995}. A handful of Be stars in the Small Magellanic Cloud undergo nova events at super soft X-ray energies \citep{sturm2012,li2012}. This provides strong support to the existence of accreting Be + white dwarf binaries, which according to binary evolution models should outnumber the Be + neutron star systems by an order of magnitude \citep{ragu2001}. Alternatively, based on correlated optical and X-ray variability \citep{robinson2002} and anti-correlation between UV and X-ray fluxes on time scales of several hours, \cite{smith1999} proposed a mechanism in which a magnetic field anchored in the rotating Be star interacts with its Keplerian circumstellar disk. Reconnection events accelerate particles that produce X-rays when impacting the disk or the photosphere \citep{robinson2000}. The origin of the putative magnetic field remains mysterious since the thick differentially rotating convective layer generating magnetic structures in solar-type stars is absent in such massive objects. 

Here we report the discovery of a strong zero delay correlation between the season averaged X-ray and optical light-curves of \gcas\ obtained over the last 14 years. X-ray data were obtained from the All-Sky Monitor (ASM) data on-board the {\it Rossi X-ray Time Explorer, (RXTE)} satellite and from the Monitor of All-sky X-ray Image (MAXI) on the {\it International Space Station}. The Automated Photometric Telescope (APT) at the Fairborn Observatory provided the optical time series. 

\section{X-ray observations}

The All-Sky Monitor \citep{levine1996} (ASM) observed $\gamma$-Cas from January 1996 until December 2011. We downloaded the ``definitive" sum band (1.5\,$-$12\,keV) light-curve accumulated over 90s dwells from the ASM RXTE database\footnote{http://xte.mit.edu/asmlc/}. All ASM data acquired in 2011 were discarded. During the last year before decommissioning a highly erratic behavior and an increase by one order of magnitude of the count rate errors prevent the use of these late data. On the average, 17 \gcas\ dwells are available every day. ASM count rates are corrected for a modeled X-ray spectrum derived from all the known sources located in the field of view. As a consequence, the count rates recorded by the ASM may become negative on occasions. For a comparatively faint X-ray source such as \gcas\ about 37\% of the sum band dwells rates are negative. Following the procedure used in \cite{trowbridge2007} and recommended by \cite{levine2013} we included negative count values in our analysis. We rebinned the 90\,s averages to longer time bins using a weighted mean process encoded in an {\it ESO-MIDAS} procedure. 

The MAXI instrument \citep{matsuoka2009} is an all-sky imager operating in the 2\,$-$20\,keV energy band installed on the Japanese Experiment Module of the {\it International Space Station} (ISS). The sky coverage is of 85\% per 92-minute orbital period and of 95\% per day. $\gamma$-Cas was observed from August 2009 on. We downloaded data averaged over one orbital period from the MAXI database\footnote{http://maxi.riken.jp/top/. Our analysis is based on data downloaded on 2014 May 7. Consequently, these data do not suffer from the incorrect storage of the flux errors reported by the MAXI team.}. Time series counts and associated errors were collected for the total 2\,$-$20\,keV band and for the 2\,$-$4\,keV, 4\,$-$10\,keV and 10\,$-$20\,keV bands. The best signal to noise ratio is achieved in the 2\,$-$20\,keV band. We rebinned the one orbit data to longer time bins using the same weighted mean process as for ASM data. MAXI light-curves contain occasional poor quality data points\footnote{see light-curve product notes at http://maxi.riken.jp/} in the form of flare-like events due to an unexpected background increase or dip-like profiles caused by shadowing by solar-battery paddles or ISS servicing spacecraft docking. In the 5 and 10\,d averages light-curves, we retained all data points with associated 1\,$\sigma$ errors below 0.01\,cts/s and count rates in the range of 0.01 to 0.05.   
We converted ASM and MAXI count rates into X-ray fluxes using {\it pimms} version 4.7 and assuming a spectral shape similar to that observed by XMM-Newton in \cite{lopes2010}, namely N$_{\rm H}$ = 2\,$\times$\,10$^{21}$\,cm$^{-2}$, abundance 0.1 solar and thermal plasma emission with $kT$\,=\,12\,keV. We found that a one ASM cts/s corresponds to 3.47\,$\times$\,10$^{-10}$\,\ergscm\ in the 1.5\,$-$\,12.0\,keV band while one MAXI broad band cts/s/cm$^{2}$ corresponds to 7.82\,$\times$\,10$^{-9}$\,\ergscm\ over the same energy band.

\begin{figure}
\includegraphics[bb= 0 0 570 650,clip=true,height=\columnwidth,angle=-90]{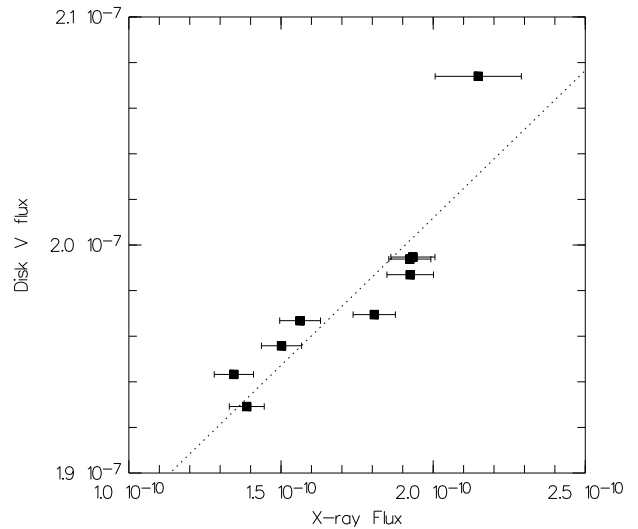}
\caption[]{The relation between the ASM (1.5\,$-$12.0\,keV) X-ray flux and the optical V band flux expected from the decretion disk (see text) in the time interval 2003-2011. These data show seasonal (yearly) averages, as discussed in the text. The dotted line shows the best fit linear relation. Fluxes are given in \ergscm. Error bars show statistical uncertainties at the one $\sigma$ level.}
\label{fluxXV}
\end{figure}

The combined high sensitivity and large field of view offered by recent all-sky X-ray monitors has yielded a wealth of high quality information on the variability of cosmic X-ray sources on time scales ranging from several years down to as short as a few seconds. However, their high ``grasp'' comes at the expense of data reduction simplicity and on occasions, can be at the origin of spurious features. For instance, coded mask instruments such as the ASM are subject to background modeling uncertainties due to the varying merged contribution of X-ray sources in the field of view. Likewise, slit + collimator instruments such as the MAXI SSC and GSC cameras may also display background and signal to noise variations due to the orbital precession of the ISS \citep{mukai2014}. We show in the appendix that the ASM and MAXI light curves of \gcas\ display periodicities of $\approx$\,200\,d and $\approx$\,70\,d, respectively, and that these periodicities are intrinsic to the all-sky monitors. These two periods are very unfortunately close to the orbital period of \gcas\ and to the mean period of the quasi-periodic oscillations seen in the optical photometry. We can nevertheless conclude, based on the ASM light curve, that 70\,d X-ray quasi-periodic oscillations are likely present and, based on the MAXI observations, that the X-ray emission is not modulated at the 203.6\,d orbital period of \gcas, in agreement with the results of \cite{shrader2015}. 

\section{Optical observations}

The APT optical monitoring consists of 16 observing seasons covering the time interval 1997 September 27 until 2013 October 2. Data are available on-line from \cite{henry2012} and from \cite{henry2014}. However, due to a calibration uncertainty, the first season cannot be used for the purpose of long term monitoring. The Johnson $B$ and $V$ filter observations of \gcas\ were made with the T3 0.4m APT at Fairborn Observatory in southern Arizona.  Full details concerning the observations can be found in \cite{henry2012} and papers listed therein. Briefly, the APT was programmed to observe \gcas\ once every two hours during every clear night of the observing season (consisting roughly of the month of June, interrupted by the Arizona summer monsoon, and September-February). In addition, several nights were targeted for concentrated
observations.  In these cases the observing sequence in both filters (with a check star and comparison star to guard against systematic errors) occurred at a cadence of 8 minutes. The {\it rms} errors are $\pm\,(0.003-0.004)$\,mag. The yearly-mean check minus comparison differential magnitudes scatter around their overall mean with an {\it rms} of 0.0019 mag. 

\section{X-ray/optical correlations}

Optical emission lines became extremely weak or even absent in the early 40's \citep{cowley1968}. At around the same time, \gcas\ reached the faintest V magnitude historically recorded at $V$\,$\sim$\,2.8 \citep{doazan1983}. It is thus likely that the star had almost completely lost its circumstellar material circa 1942 and that the V magnitude recorded at that time represents the brightness of the naked star. Over the time interval 1998$-$2013 covered by APT monitoring, the average optical emission of the circumstellar disk had increased the brightness of \gcas\ by about 0.6 magnitude in V with respect to naked stellar emission. Accordingly, we estimated disk optical emission by subtracting from the APT V flux that observed from the naked star. V band fluxes were computed assuming an effective filter width of 880\,\AA, and a V zero point of 3.75\,$\times$\,10$^{-9}$ erg\, s$^{-1}$\,cm$^{-2}$\,\AA$^{-1}$ \citep{allen2001}. The V band disk flux is then given by: \begin{equation}
F_{\mathrm V}^{\mathrm{Disk}} = 3.75\,10^{-9}\,\times\, 880\,\times\,(10^{-\frac{\mathrm V}{2.5}} - 10^{-\frac{\mathrm 2.8}{2.5}}) \ \mathrm{erg}\,\mathrm{s}^{-1}\,\mathrm{cm}^{-2}
\label{fvfx}
\end{equation}

\begin{figure}
\includegraphics[bb= 50 40 570 790,clip=true,height=\columnwidth,angle=-90]{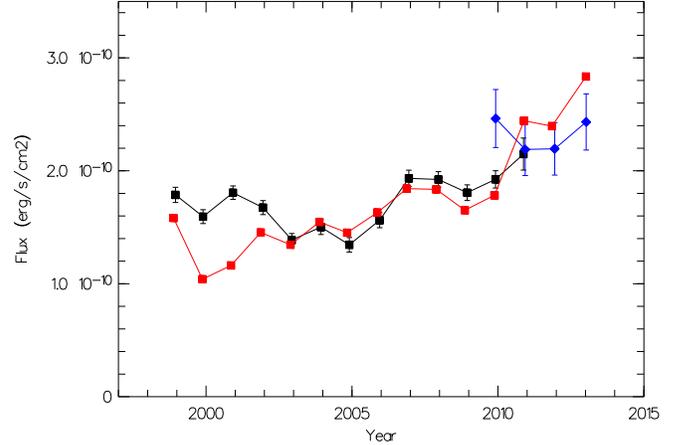}
\caption[]{ASM, MAXI and V band from APT monitoring yearly season light-curves. 
Black squares: ASM (1.5\,$-$\,12.0\,keV) X-ray flux. Blue diamonds: MAXI fluxes in the ASM energy range. The spurious $\approx$\,200\,d ASM modulation has been removed. Red squares: V band flux from the decretion disk scaled down to the X-ray flux range using the relation derived from ASM and V band data after 2002 (see text). MAXI data follow the same trend as earlier ASM data after 2010. The larger MAXI flux error bars account for the 10\% cross calibration uncertainty, however, their intrinsic mean error is of only 7\,$\times$\,10$^{-12}$\,\ergscm. The origin of the discrepancy between ASM and MAXI fluxes in 2009 is unknown and since the ASM and MAXI fluxes are identical in 2010, we suspect a calibration problem during the first MAXI operation year. Error bars show statistical uncertainties at the one $\sigma$ level. Errors on scaled down V flux are too small ($\sim$ 10$^{-12}$\ergscm) to be visible.}
\label{XrayVseasons}
\end{figure}

\begin{figure}
\begin{center}
\includegraphics[bb= 60 65 930 745,clip=true,width=0.75\columnwidth,angle=0]{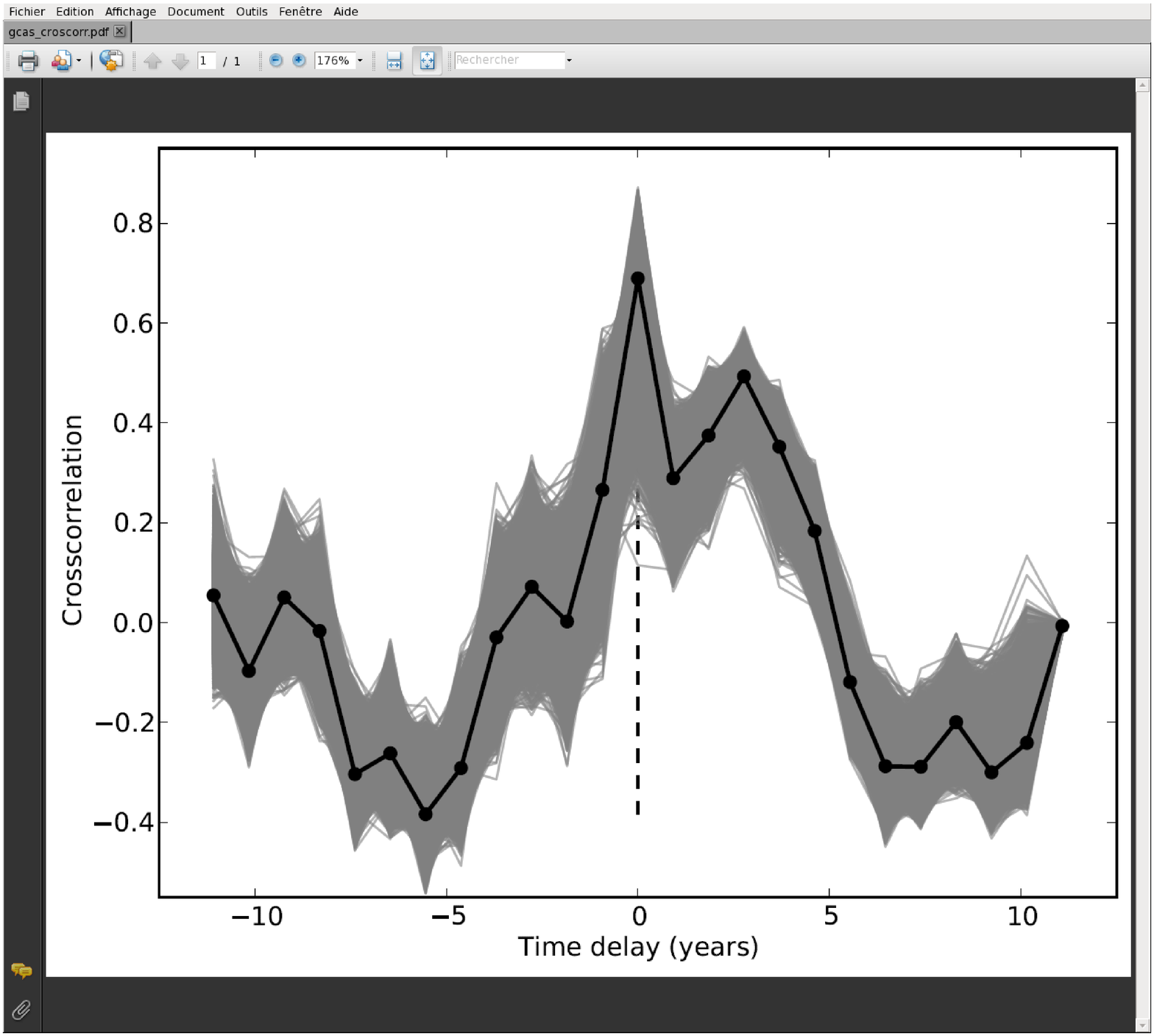}
\caption[]{The crosscorrelation function of the ASM and APT light-curves shown on Fig.~\ref{XrayVseasons}.  The gray parts correspond to the superposition of 50,000 crosscorrelations which were constructed from 50,000 pairs of fake light curves. The crosscorrelation of the original light curves is shown in black.}
\label{gcas_crosscorr}
\end{center}
\end{figure}

\begin{figure*}
\begin{center}
\includegraphics[bb= 90 60 524 791,clip=true,width=1.8\columnwidth,height=\textheight,angle=0]{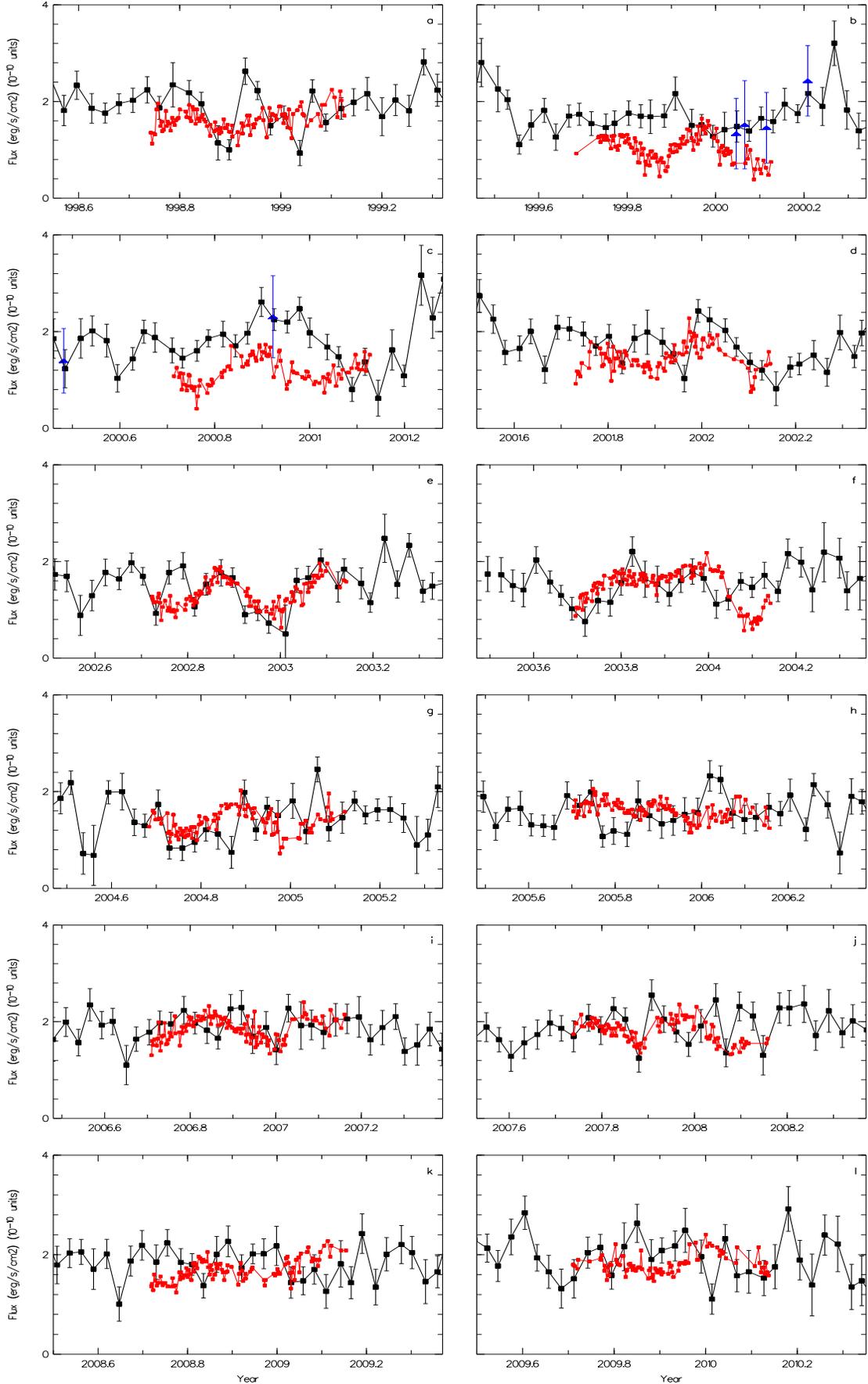}
\caption[]{ASM and V band flux light-curves. Times are in years. Black squares: ASM data (1.5\,$-$\,12\,keV) averaged over 10\,d time bins after removal of the spurious $\approx$\,200\,d modulation (see appendix). Red squares represent the 1\,d averages of the variable part of the optical circumstellar disk emission scaled down to the X-ray flux range (see text). Error bars show statistical uncertainties at the one $\sigma$ level.  Blue triangles and associated error bars mark the full range of X-ray flux detected in 2000 by the PCA on {\it RXTE} as based on 30\,minute averages \citep{robinson2002}.}
\label{ASMVseasons}
\end{center}
\end{figure*}

APT optical data are not evenly distributed in time due to the visibility of the star and to the monsoon season. In order to build long term optical and X-ray light curves, we grouped optical data inside specific windows that for each year exclude a few (about 8\%) scattered V and B magnitudes obtained before or after the main observing season which extends from approximately September to January. We computed ASM and MAXI count rate averages over these time windows so as to ensure the best simultaneity between X-ray and optical data. Over the time interval from 1998 until 2012, the mean duration of the yearly time bin is 0.43\,yr. However, because of the much coarser coverage obtained during the last optical runs, the last data point is averaged over a 1.34\,yr long time interval from May 2012 until October 2013. Fig.~\ref{fluxXV} shows that a well defined correlation between X-ray and disk optical emission exists on time scales of 9 years from 2002 until 2010.  The relation between the ASM (1.5\,$-$12.0\,keV) flux and the optical V band emission from the decretion disk computed using (\ref{fvfx}) can be simply fitted by a linear relation. Taking into account errors on optical and X-ray flux we find:
\begin{equation}
F_{\mathrm V}^{\mathrm{Disk}} = 129\,(\pm 14)\,\times\,F_{\mathrm X}^{\mathrm{ASM}}\ + 1.75\,(\pm 0.02)\,10^{-7} \  \mathrm{erg}\,\mathrm{s}^{-1}\,\mathrm{cm}^{-2}
\label{linear}
\end{equation}
Extrapolating this relation at null X-ray flux and assuming that a linear regime holds over the whole flux range implies a ``basal" V band flux equivalent to 85\,($\pm$\,6)\% of the disk V band flux observed in late 2010. Here we assume an error of $\pm$ 0.05 mag on the V magnitude of the naked star. This suggests that only a relatively small fraction of the regions of the decretion disk contributing to V band emission is related to the production of X-rays. Our $\Delta F_{\mathrm V}^{\mathrm{Disk}}/\Delta F_{\mathrm X}^{\mathrm{ASM}}$ ratio of $\approx$\,130 is in agreement with the value of $\approx$\,80 found by \cite{smith2006} when correcting for stellar contribution. 
We then used this relation to scale the optical flux of the decretion disk to the X-ray range and plot on the same frame the ASM, MAXI and APT season averages shown in Fig.~\ref{XrayVseasons}.  The light curves confirm the amazingly good correlation between X-ray and optical data acquired after 2002, extending to the time of the MAXI observations. A similar correlation probably occurred before 2002, albeit with a slightly shifted zero point.  Brighter optical disk emission is invariably accompanied by a similar increase of the X-ray flux. 

We estimated the maximum lag between optical and ASM flux variations using a Monte Carlo method simulating a total of 50,000 different light-curves with the same statistical properties as the observational ones. The crosscorrelation function of the ASM and APT light curves and the results of the Monte Carlo simulations are shown in Fig.~\ref{gcas_crosscorr}. We find that any delay between X-ray and V band flux variations is less than about one month at the 3\,$\sigma$ level. 

\begin{figure}
\begin{center}
\includegraphics[bb= 40 40 500 500,clip=true,width=\columnwidth,angle=0]{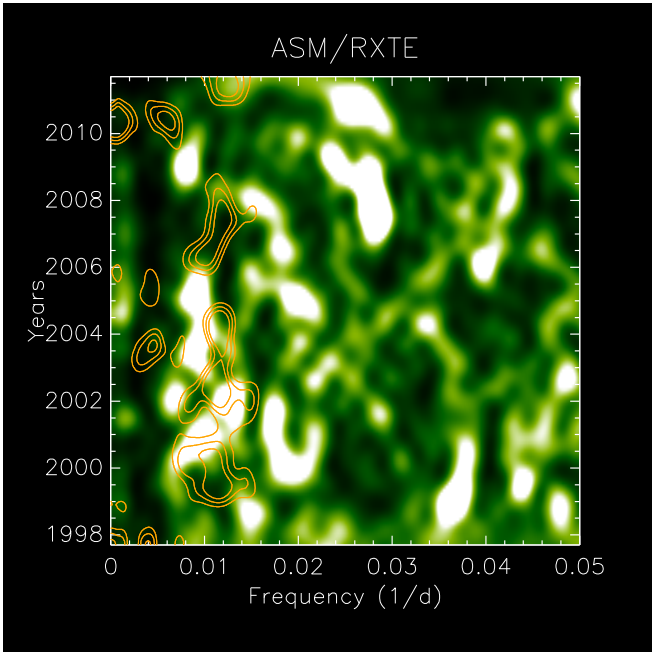}
\caption[]{X-ray and optical periodograms. The periodogram of the optical V band data (orange contours showing normalized powers 15,20 \& 25) is overplotted on the ASM (1.5\,$-$\,12.0\,keV) periodograms (normalized power range from 0 to 2). All periodograms are computed on a one year long interval sliding by 0.1 year and Gaussian filtered in the time direction with a sigma of 0.3 year. ASM data have a 5 day time resolution while optical measurements are averaged over one day. ASM data have been cleaned from the spurious 200\,d artifact. Some excess power is detected in the period range from 50\,d to 100\,d. The strongest optical $\approx$\,80\,d oscillations are weakly correlated with the largest X-ray power.}
\label{ASMVper}
\end{center}
\end{figure}

\begin{figure}
%\centerline{\psfig{file=xplotQPOsFromAnyStrip.ps,bbllx=40pt,bblly=85pt,bburx=570pt,bbury=790pt,clip=true,width=8cm,angle=-90}}
\begin{center}
\includegraphics[bb= 40 85 570 790,clip=true,height=\columnwidth,angle=-90]{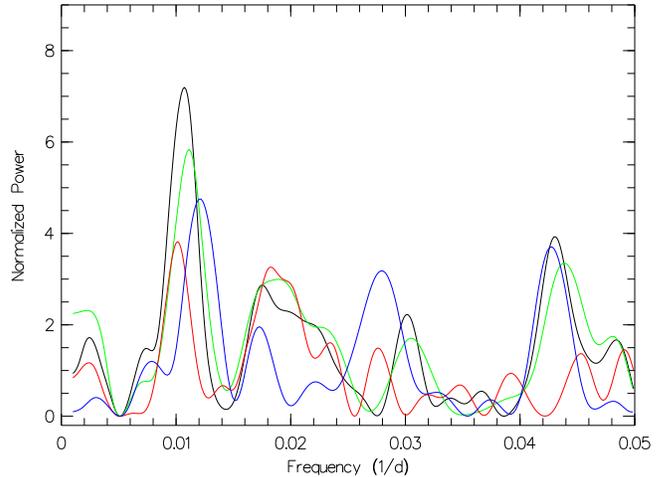}
\caption[]{Power spectra of strips of ASM data. Black:  2000.8 - 2001.8, green: 2001 - 2001.8, red: 2005.2 - 2006.2 and blue; 2008.5 - 2009.2. QPOs with periodicities in the range of 70\,d to 110\,d are detected in these particular time intervals. However, QPO periods down to about 50\,d are also seen on occasion (see Fig. \ref{ASMVper}).}
\label{QPOpower}
\end{center}
\end{figure}

The binned ASM X-ray light-curves suggest the presence of oscillations on time scales of $\approx$ 50 to 100\,d (see Fig.~\ref{ASMVseasons} a,e for instance and  Fig.~\ref{ASMVper}). Similar time scales have been seen in the optical time series collected since 1996 \citep{smith2006}; see also Figs.~\ref{ASMVseasons} and \ref{ASMVper}. Overall, the statistical significance of the X-ray QPOs is not very high. However, X-ray QPOs exhibit stronger amplitudes during some time intervals. The four power spectra shown in Fig.~\ref{QPOpower} have bin probabilities $\leq$\,2\% to be due to a white noise random fluctuation at the optical QPO periods. It seems therefore very likely that the X-ray light curve of \gcas\ is also modulated with quasi-periodic oscillations at a mean period very close to that regularly detected in the optical photometry. For example, \cite{robinson2002} reported evidence of a $\approx$ 70\,d X-ray modulation in a series of 6 snapshot {\it RXTE}/PCA observations obtained in 2000. Accordingly, we overplot on Fig.~\ref{ASMVseasons}~{b \& c} the scaled observed PCA flux on the top of the equivalent ASM 10\,d broad band averages. Error bars represent the total range of PCA count rates averaged over 30\,mn time intervals observed during each of these $\approx$ 27\,h continuous {\it RXTE}/PCA observations \citep[][Fig.~6]{robinson2002}. Mid PCA fluxes match very well those derived from ASM data providing an independent confirmation of the reliability of the ASM light curves. Although in 2000, the 70\,d QPOs are detected at both X-ray and optical energies, with hardly any time delay, it can be seen that the relatively few {\it RXTE}/PCA short observations cannot accurately sample the full complexity of the X-ray light curve.

In spite of the very likely existence of a $\approx$\,70\,d signal at both X-ray and optical wavelengths, the overall crosscorrelation between ASM X-ray fluxes and optical fluxes does not show any significant peak. This probably reflects the complex and changing relation patterns between optical and X-ray flux variations on weekly timescales. However, on several occasions strong and well correlated trains of optical and X-ray modulations are clearly detected with small absolute time lags (see e.g. Fig.~\ref{ASMVseasons}~{c, d, e \& i}).  

Importantly, the relative amplitudes of the X-ray and optical modulations on time scales of a few weeks are the same as those derived from yearly averages, thus hinting at a common mechanism for both the long and short time scales variations.

\section{Discussion and Conclusions}

The optical brightness of \gcas\ ($V\,\approx\,2.4$) has turned the star into an obvious target for interferometric campaigns. Observations carried out in the visible (\Halp\ continuum) and near infrared (H band) between 2010 Aug. 5 and Nov. 10 \citep{stee2012} give a disk diameter of $\approx$ 1.9\,\Dstar\ indicating that most of the visible and near-IR disk emission arises from the 0.45\,\Rstar\ wide inner part of the disk above the photosphere. This is consistent with viscous decretion disks modeling \citep{carciofi2006,carciofi2011} which indeed predict that for systems similar to \gcas, V band emission is confined to within 1\,\Rstar\ from the B star. Therefore, the $\approx$\,70\,d V band QPOs as well as the long term changes must originate from relatively close to the star. The mechanism giving rise to the 70\,d QPOs is not known, although the possibility that these oscillations are due to a dynamo operating in the decretion disk has been proposed by \cite{robinson2002}. At any rate, the 70\,d periodicity is probably not directly related to Keplerian rotation. Since \gcas\ is rotating at or nearly at the critical equatorial velocity \citep{stee2012} with a period of $\approx$\,1.22\,d \citep{henry2012}, a 70\,d Keplerian rotation period is reached at R = (70/1.22)$^{2/3}$\,\Rstar\ $\approx$ 15\,\Rstar, well beyond the end of the V band emitting region. The proximity of the ratio between orbital and mean QPO periods to 3:1 suggests a possible resonance effect. Indeed, low eccentricity systems such as \gcas\ are expected to display moderate disk truncation at the 3:1 resonance radius \citep{okazaki2001}. Models of tidally perturbed disks show density spiral patterns that may extend to the inner parts of the disk. Such an orbitally ``locked" mechanism would explain why the 70\,d QPOs clock is at work independently of disk optical brightness \citep{smith2006}. Observational evidence in favor of tidally induced effects remain rather scarce. One example is the binary Be star $\pi$ Aqr, the \Halp\ profile of which shows orbital phase-locked Violet/Red lobe variations apparently caused by a disk region of enhanced density facing the companion star \citep{zharikov2013}. 

Our analysis reveals for the first time the excellent long term correlation between the yearly averages of the V band brightness of the circumstellar disk and X-ray emission. X-ray and optical correlated behaviors of the 70\,d QPOs were suspected by \cite{robinson2002} based on a coarse X-ray monitoring. We show here that although such a strong correlation may happen on occasion, the actual correlation pattern is complex. 

The high value of the ratio between optical and X-ray flux variations on time scales of years or on time scales of the order of the 70\,d QPOs, ($\Delta F_{\mathrm V}^{\mathrm{Disk}}/\Delta F_{\mathrm X}^{\mathrm{ASM}}$\, $\approx$\,130) rules out a mechanism in which optical light changes are due to X-ray heating. A similar conclusion has been reached by \cite{smith1998a}.  In principle this does not exclude the possibility that a mass ejection event first detected as a surge in the V band from the inner disk migrates to the orbit of the companion star, is accreted and gives rise to an increase of X-ray luminosity. Likewise, the density modulations in the inner part of the decretion disk seen as 70\,d optical QPOs could perhaps be transported to a putative accreting companion and modulate the accretion rate on similar time scales. However, the near simultaneity of the long term optical and X-ray variations rules out this possibility. 

%\begin{figure}[ht]
\begin{figure}
\begin{center}
\includegraphics[bb= 40 75 575 800,clip=true,height=\columnwidth,angle=-90]{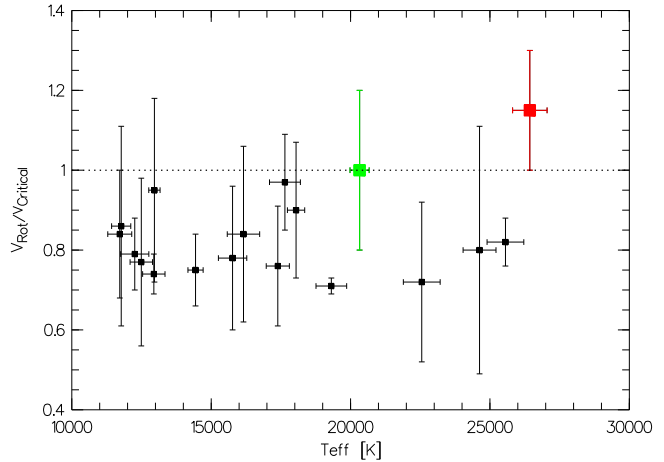}
\caption[]{The distribution of Be stars in the \Teff\ versus V$_{\mathrm{rot}}$/V$_{\mathrm c}$ diagram for stars having inclinations measured by long base interferometric observations. Green symbol: HD 110432. Red symbol: \gcas.}
\label{Wteff}
\end{center}
\end{figure}

\begin{table*}[ht]
\begin{minipage}{\textwidth}
\begin{center}
\caption{X-ray (ROSAT) and interferometric rotational parameters of the Be stars plotted in Fig.~\ref{Wteff}. \newline}
\begin{tabular}{rrclccc}
\hline
\hline
  HD      &Log(Lx)& V$_{\mathrm{rot}}$/V$_{\mathrm c}$& Spectral &  \Teff      &  i   & Ref.\footnote{{\textbf{ References.}} (1) \cite{berghoefer1996}; (2) \cite{cohen1997}}\\   
 number    & (0.1-2.4\,keV) &                        & type     &     (K)       & (deg)  & \\ \hline
   5394\footnote{\ HD\,5394 = \gcas} &	 32.13 &  1.15 $\pm$ 0.15 & B0.5IVe  &  26431 $\pm$  618  &  47  & 1\\
  10516 & $\leq$ 29.78 &  0.82 $\pm$ 0.06 & B1.5Ve   &  25556 $\pm$  659  &  72  & 1\\
  50013 &	 29.85 &  0.80 $\pm$ 0.31 & B1.5V    &  24627 $\pm$  590  &  31  & 2\\
 120324 &	 30.45 &  0.72 $\pm$ 0.20 & B2IV     &  22554 $\pm$  661  &  18  & 2\\
 110432 &	 31.08\footnote{\ \Lx\ = 4.2-5.2\,$\times$\,10$^{32}$\ergs\, (0.2-12\,keV) \citep{lopes2007}}  &  1.00 $\pm$ 0.20 & B0.5-1   &  20324 $\pm$  344  &  63  & 1\\
  37202 & $\leq$ 29.27 &  0.71 $\pm$ 0.02 & B2IVp    &  19310 $\pm$  550  &  78  & 1\\
 158427 &	 28.57 &  0.90 $\pm$ 0.17 & B3IV     &  18044 $\pm$  310  &  39  & 2\\
 142983 & $\leq$ 30.02 &  0.97 $\pm$ 0.12 & B8Ia/Iab &  17645 $\pm$  554  &  57  & 1\\
  91465 & $\leq$ 29.15 &  0.76 $\pm$ 0.15 & B4V      &  17389 $\pm$  415  &  45  & 1\\
  25940 & $\leq$ 29.36 &  0.84 $\pm$ 0.22 & B3Ve     &  16158 $\pm$  582  &  43  & 1\\
  22192 & $\leq$ 29.27 &  0.78 $\pm$ 0.18 & B5Ve     &  15767 $\pm$  509  &  72  & 1\\
   4180 & $\leq$ 29.83 &  0.75 $\pm$ 0.09 & B5IIe    &  14438 $\pm$  272  &  57  & 1\\
  37795 &	 27.75 &  0.95 $\pm$ 0.23 & B7IV     &  12963 $\pm$  203  &  33  & 2\\
 209409 & $\leq$ 29.49 &  0.74 $\pm$ 0.05 & B7VIe    &  12942 $\pm$  402  &  78  & 1\\
 209409 & $\leq$ 29.49 &  0.77 $\pm$ 0.21 & B7IV     &  12492 $\pm$  402  &  45  & 1\\
  23630 & $\leq$ 28.98 &  0.79 $\pm$ 0.09 & B7IIIe   &  12258 $\pm$  505  &  44  & 1\\
  58715 & $\leq$ 28.49 &  0.86 $\pm$ 0.25 & B8Ve     &  11772 $\pm$  344  &  45  & 1\\
  89080 & $\leq$ 28.87 &  0.84 $\pm$ 0.16 & B8III    &  11720 $\pm$  431  &  55  & 1\\
\hline
\end{tabular}
\label{tlx}
\end{center}
\end{minipage}
\end{table*}

Viscous disks response to sustained mass injection has been investigated by \cite{okazaki2002} and \cite{carciofi2012} among others. Dynamical models by \cite{haubois2012} show that the inner parts of the disk are first filled in and that at radii equivalent to that of the primary Roche lobe ($R\,\approx\,$21\Rstar\ for \gcas\ \citep{gies2007}), steady state disk densities are only reached after typically 5 to 20 years depending on the assumed $\alpha$ viscosity. The dissipation of the decretion disk of the Be star 28 CMa is well described by these models \citep{carciofi2012}. Unfortunately, to our knowledge, only few Be/X-ray systems have long enough X-ray and optical light curves to test the validity of the dynamical models at relatively large distances of the B star. The first example is SXP18.3, a Be/X-ray neutron star binary in the SMC with a 17.79\,d orbital period which experienced a period of strong X-ray activity about 4 years after the beginning of a major mass ejection event seen optically \citep{schurch2009}. Nearly the same delay between major optical and X-ray outbursts may have been seen in the low \Lx\ persistent X-ray pulsar X~Per \citep{li2014}. With an orbital period of $\approx$\,250\,d, an eccentricity of $e$\,=\,0.11 and an O9.5IIIe-B0Ve primary, X~Per resembles \gcas. This system does not exhibit strong X-ray flux variations with periastron passage. However, no significant correlation with time lags lower than a few years is seen between optical and X-ray light curves \citep{lutovinov2012}, at variance with the observed behaviour of \gcas. Finally, the Be/X-ray binary A\,0535+26 terminated in 2005 its ten year long inactivity period by a series of giant and normal outbursts starting about five years after the onset of a V band flux increase marking the rebuilding of a new decretion disk \citep{yan2012,camero2012}. 
In these systems, the long time delays between outbursts in X-rays, due to the sudden availability of matter for accretion at the neutron star orbital distance, and in optical signatures of a sustained increase of the decretion density close to the star, are consistent with the predictions of the dynamical models. Consequently, the maximum one month lag between X-ray and optical long term changes completely rules out accretion on a compact object in a \,203.6\,d orbit as the X-ray emission mechanism in \gcas. Optical variations can be easily understood as changes in the emission measure in regions within the inner disk. These arguments are consistent with magnetic disk-star interactions as the most natural X-ray emitting mechanism, as originally proposed \citep{smith1999}. We develop this point next.

The ``Magnetism in Massive Stars" (MiMes) project \citep{grundhut2012} showed that strong and structured magnetic fields of likely fossil origin can be found in about 7\% of all massive O and B stars. These fossil fields may have their origin in a convecting phase occurring during pre-main sequence stage, in an early binary merger phase, or be remnant of a primordial field \citep[see][for a recent review]{walder2012}. However, one of the most intriguing result of MiMes is the total lack of magnetic field detection in classical Be stars \citep{grundhut2012} with upper limits of B\,$\sim$\,250\,G and B\,$\sim$\,500\,G for organized and small scale fields with filling factors of 50\% respectively \citep{rivinius2013}. Importantly, \gcas\ also fails to show evidence of global magnetism with similar upper limits \citep{wade2014}, although the detection of the stellar rotation period in the optical photometry \citep{henry2012} suggests the presence of some magnetic spots at the surface of the B star. A magnetized wind can efficiently carry away angular momentum and spin down stellar rotation with time scales of the order of a few Myr \citep{petit2013}, significantly shorter than the estimated ages of \gcas\ and of its analogs \citep{smith2012}. The apparent absence of strong and structured magnetic field in Be stars is therefore consistent with their high rotation rate.

Thin subsurface convecting layers may exist in all massive stars as a result of the iron peak in stellar opacities \citep{cantiello2009}. The thickness and mass content of the convective layers increases very substantially and are more equatorially condensed for stars rotating close to the critical velocity \citep{maeder2008} such as \gcas. This mechanism could explain the presence of strong, localized magnetic fields connecting the stellar equator to the decretion disk and generating the anomalous X-ray emission. Being mostly located in the equatorial plane such random magnetic fields may remain concealed from optical spectro-polarimetry observations. 

Long baseline interferometric survey observations of Be stars have been recently reported by \cite{meilland2012} and \cite{touhami2013}. Disk axial ratios provide information on inclination and together with observed velocities, suggest that on the average, Be stars rotate at about 0.8 times the critical velocity. Similar conclusions have been reached by \cite{fremat2005} based on the modeling of optical spectra taking into account the flattening and varying gravitational darkening caused by the fast rotation. 
Interestingly, long baseline optical interferometric observations of the the two brightest \gcas\ analogs, \gcas\ itself (B0.5 IV-Ve) and HD\,110432 (B0.5 IIIe) confirm that they are both nearly critical rotators (V$_{\mathrm{rot}}$/V$_{\mathrm c}$ = 1.15\,$\pm$\,0.15 and V$_{\mathrm{rot}}$/V$_{\mathrm c}$ = 1.0\,$\pm$\,0.20 respectively \citep{stee2012,stee2013}). We note that for these two stars, the inclinations computed by \cite{fremat2005} are substantially higher than those derived from long base interferometric observations, suggesting a possible breakdown of the optical spectral modeling at close to critical velocities. 

We show in Fig.~\ref{Wteff} the distribution of 18 Be stars with interferometrically measured inclinations in the \Teff\ versus V$_{\mathrm{rot}}$/V$_{\mathrm c}$ diagram. We excluded the ill-defined $\delta$ Cen value \citep{meilland2012}. All \Teff\ were taken from \cite{fremat2005}. All Be stars marked in Fig.~\ref{Wteff} have been either detected in X-rays or have significant upper limits on their X-ray luminosity, either from the ROSAT all-sky survey \citep{berghoefer1996,voges1999}, or from the ROSAT survey of Be stars conducted by \cite{cohen1997}. The corresponding values are listed in Table \ref{tlx}.

Although error bars remain large, Fig.~\ref{Wteff} shows that both \gcas\ and its analog HD\,110432 are located in a region of extreme rotation rates and hot effective temperatures apparently void of normal Be stars. In particular, none of the four highest \Teff\ ($\geq$ 19,000 K) Be stars with V$_{\mathrm{rot}}$/V$_{\mathrm c}$ $\leq$ 0.9 (HD 10516 = $\phi$ Per, HD 50013 = $\kappa$ CMa, HD 120324 = $\mu$ Cen and HD 37202 = $\zeta$ Tau ) exhibit an X-ray luminosity in the typical range of \gcas\ analogs\footnote{XMM-Newton detection parameters and spectra of HD 120324 were extracted from the XCat-DB (http://xcatdb.unistra.fr/3xmm/). The observed XMM-Newton X-ray flux yields a revised Log(Lx) (0.1$-$10\,keV) of 29.9. The EPIC pn X-ray spectrum is soft and requires at least three thermal components with a maximum temperature of $\approx$ 2\,keV.}.  

We therefore propose that \gcas-like objects could represent {\em the most rapidly rotating of the most massive Be stars in which magnetic field generation is the direct consequence of critical rotation}. The fact that some of the \gcas\ analogs are blue stragglers \citep{marco2009,smith2012} which could have been spun up to highest rates through binary evolution provides further support to this scenario.  
However, the very narrow range of B0e-B1.5e spectral types in which the \gcas\ phenomenon is observed \citep{smith2015} certainly calls for an explanation. The hot boundary is likely related to the disappearance of circumstellar disks at this spectral type due to radiative pressure or due to a different evolution of the angular momentum \citep[see e.g.][and references therein]{negueruela2004}. The lower mass limit may reflect the steep decrease of the size of the convective layers around B2 and may be related to the observed sharp drop of the \Lx/\Lbol\ ratio at about the same spectral type \citep{cohen1997}.  

Long base interferometry of fainter \gcas-analogs with V magnitudes of $\approx$\,6.6 or more \citep{smith2015} using the most advanced instruments currently available (e.g. CHARA and the ESO-VLTI) may allow us to measure the inclination of their circumstellar disk with enough accuracy so as to constrain their equatorial rotation rate and test the validity of this scenario. Conversely, a systematic X-ray survey of the fastest rotating early Be stars may help to precise the critical rotation rate and effective temperature above which the \gcas\ phenomenon starts and constrain the mechanism giving rise to the putative magnetic field in \gcas-analogs. 

\acknowledgments
We thank the anonymous referee whose comments greatly improved the presentation and clarity of the paper. R.L. was supported by the Brazilian agency CNPq (Universal Grants 470361/2011-5). R.L. thanks Evandro M. Ribeiro for help in coding the crosscorrelation analysis. R.L. and C.M. acknowledge support from CNPq APV Grants 454852/2012-6. We are greatful to Koji Mukai for providing insights on the effect of ISS orbital precession on MAXI light curves and to G. Henry for giving us access to unpublished 2013 APT data. Results are provided by the ASM/RXTE teams at MIT and at the RXTE SOF and GOF at NASA's GSFC. This research has made use of MAXI data provided by RIKEN, JAXA and the MAXI team. 

{\it Facilities:} \facility{RXTE (ASM)}, \facility{MAXI}, \facility{APT at Fairborn Observatory}.

\clearpage 

\bibliography{ms_gcas.bib}

\bibliographystyle{apj}



\clearpage

\appendix

\section{Periodicities intrinsic to the ASM and MAXI detectors}

In this appendix, we discuss the reality of some of the signals delivered by ASM/RXTE and MAXI at periods  close to those intrinsically displayed by \gcas, namely, $\approx$ 200\,d (orbital period) and $\approx$ 70\,d (optical quasi periodic variations). All timing analysis was performed using the least square power spectrum (LSPS) \citep{lomb1976,scargle1982} which is well suited to the unevenly distributed optical data while providing good results on the more regularly sequenced X-ray time series.

\subsection{A $\sim$\,200\,d ASM modulation artifact}

\begin{figure}
\begin{center}
\includegraphics[bb= 40 70 570 790,clip=true,height=\columnwidth,angle=-90]{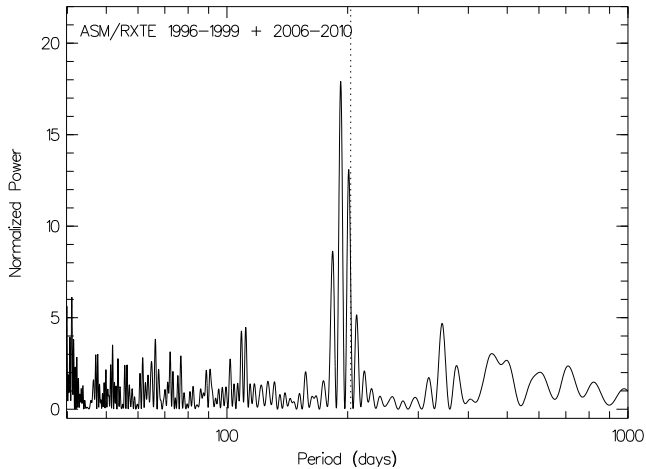}
\caption[]{The Lomb-Scargle power spectrum all ASM/RXTE broad band data grouped in 10\,d bins of time. The two higher peaks are at 192\,d and 201\,d. The dashed line shows the position of the orbital period of \gcas.}
\label{powerspec200}
\end{center}
\end{figure}

The LSPS of the entire ASM 1996-2010 broad band time series shows a group of peaks around 200\,d tantalizingly close to the 203.6\,d orbital period of \gcas. ASM power spectra in the time intervals 1996-1999 and 2006-2010 (dates inclusive) exhibit clear modulations at $\approx$\,193\,d, while no such peak is present in between. A LSPS applied to the merging of the two time intervals listed above yields a main peak at 192\,d with a weaker 201\,d alias (see Fig.~\ref{powerspec200}). With a probability of 4\,$\times$\,$10^{-6}$ to be detected at random over the 20\,d to 1000\,d range of periods searched, the $\sim$ 192\,d signal is highly significant with full amplitudes in the range of 0.10 to 0.14 cts/s. Unfortunately, this periodicity is also close to half a year and should therefore be considered with caution. Although the window spectrum does not show evidence of a signal at these periods, the background level does show a clear annual variation. Taken separately, both negative and positive rate time series display considerable power at a period of exactly six months (182.6\,d). It seems thus clear that the actual background used to correct the light-curves varies with a half year period and that this modulation could impact our ability to detect a genuine astrophysical signal close to the 203.6\,d orbital period of \gcas. However, the reason that the detected period is slightly longer than 6 months is unclear and may be due to a beating phenomenon between the half year period and a longer $\sim$ 10\,yr modulation. \cite{wen2006} have investigated in great details the overall periodicities found in ASM light-curves and identified a number of likely spurious frequencies. In particular, periods in the range of 163\,d to 196\,d are detected in 17 sources and were therefore considered as likely artifacts. Using power spectra of AGN or supernova remnant, \cite{levine2011} show that spurious signals may be found in the period range of 174.5\,d to 191.2\,d. Therefore, the occurrence of similar periodicities in the X-ray light curves of other ASM sources and the strong background modulations with a half-year period, all point at a non astrophysical origin of 192\,d modulation and of its 201\,d alias. Consequently in this paper, we use ASM light-curves corrected for the $\sim$\,200\,d periods. However, we systematically checked that using original data without corrections for the spurious 200\,d period yield results consistent with those obtained based on corrected time series. 

\subsection{A $\sim$\,70\,d MAXI modulation artifact}
 
\begin{figure}
\begin{center}
\includegraphics[bb= 40 70 570 790,clip=true,height=\columnwidth,angle=-90]{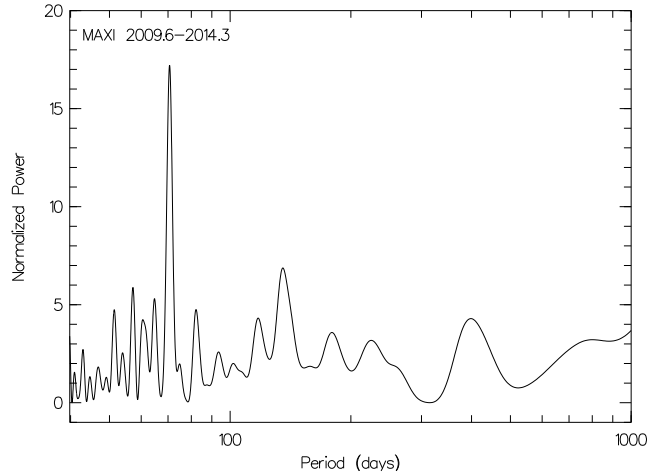}
\caption[]{The Lomb-Scargle power spectrum of \gcas \ MAXI broad band data grouped in 5\,d bins of time. The highest peak is at $\approx$ 70.5\,d, a period close to the precession period ($\approx$ 70\,d) of the orbit of the {\it International Space Station} on which the MAXI instrument is installed.}
\label{powerspecMAXI}
\end{center}
\end{figure}

\begin{figure}
\begin{center}
\includegraphics[bb= 40 70 570 790,clip=true,height=\columnwidth,angle=-90]{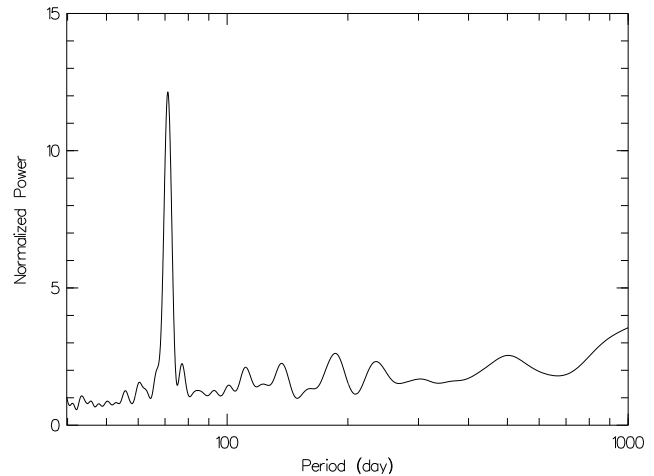}
\caption{Average Least Square Power Spectrum of the 11 MAXI sources with continuous light curves and with mean count rates similar to that of \gcas\ (0.01-0.05 cnt/s). This result can be compared to Fig.~\ref{powerspecMAXI} .}
\label{averagelsps}
\end{center}
\end{figure} 
 
\begin{center}
\begin{figure}
\begin{tabular}{c}
\includegraphics[bb= 40 65 570 620,clip=true,height=\columnwidth,width=0.31\textheight,angle=-90]{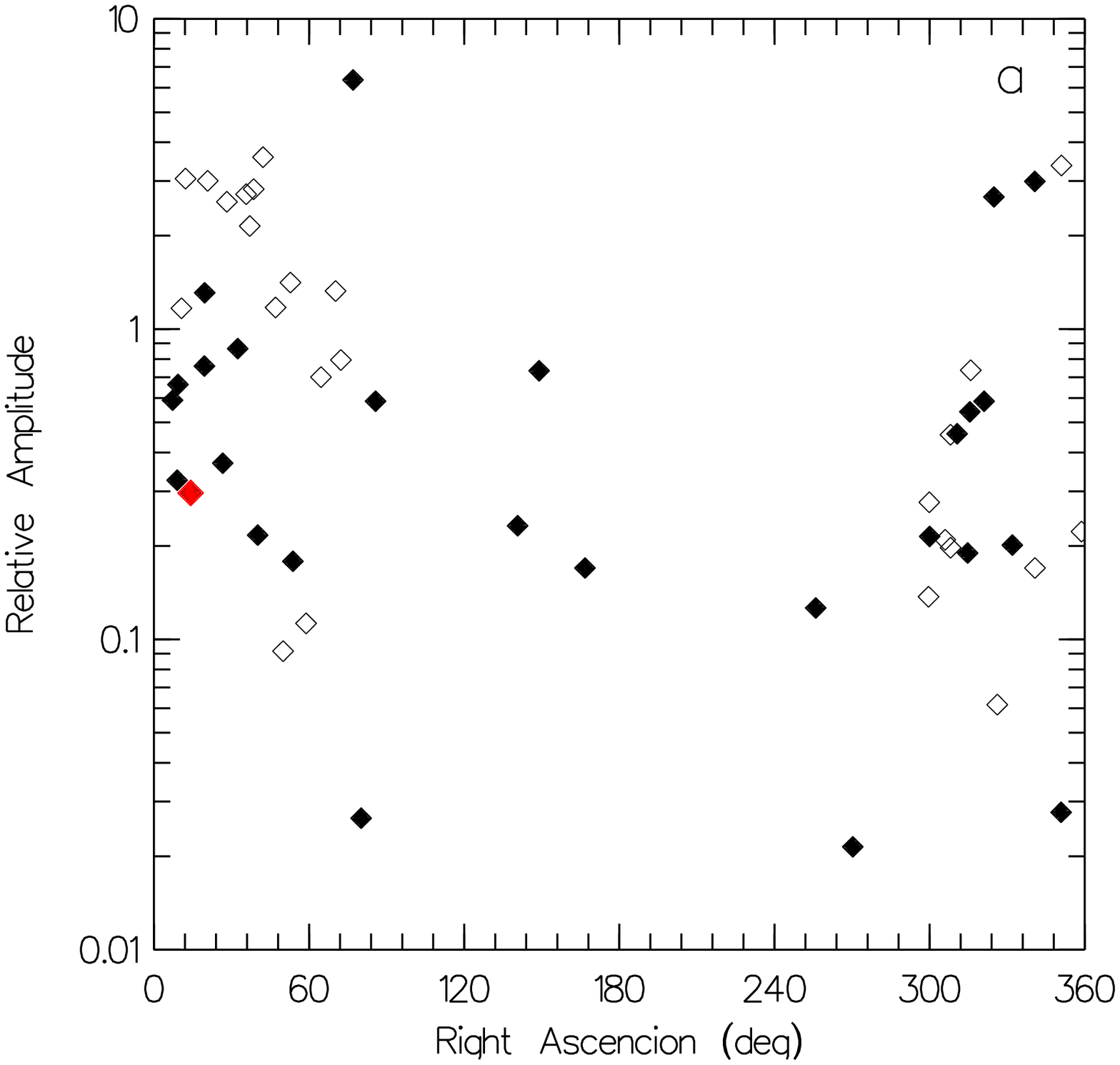} \\
\includegraphics[bb= 40 65 570 620,clip=true,height=\columnwidth,width=0.31\textheight,angle=-90]{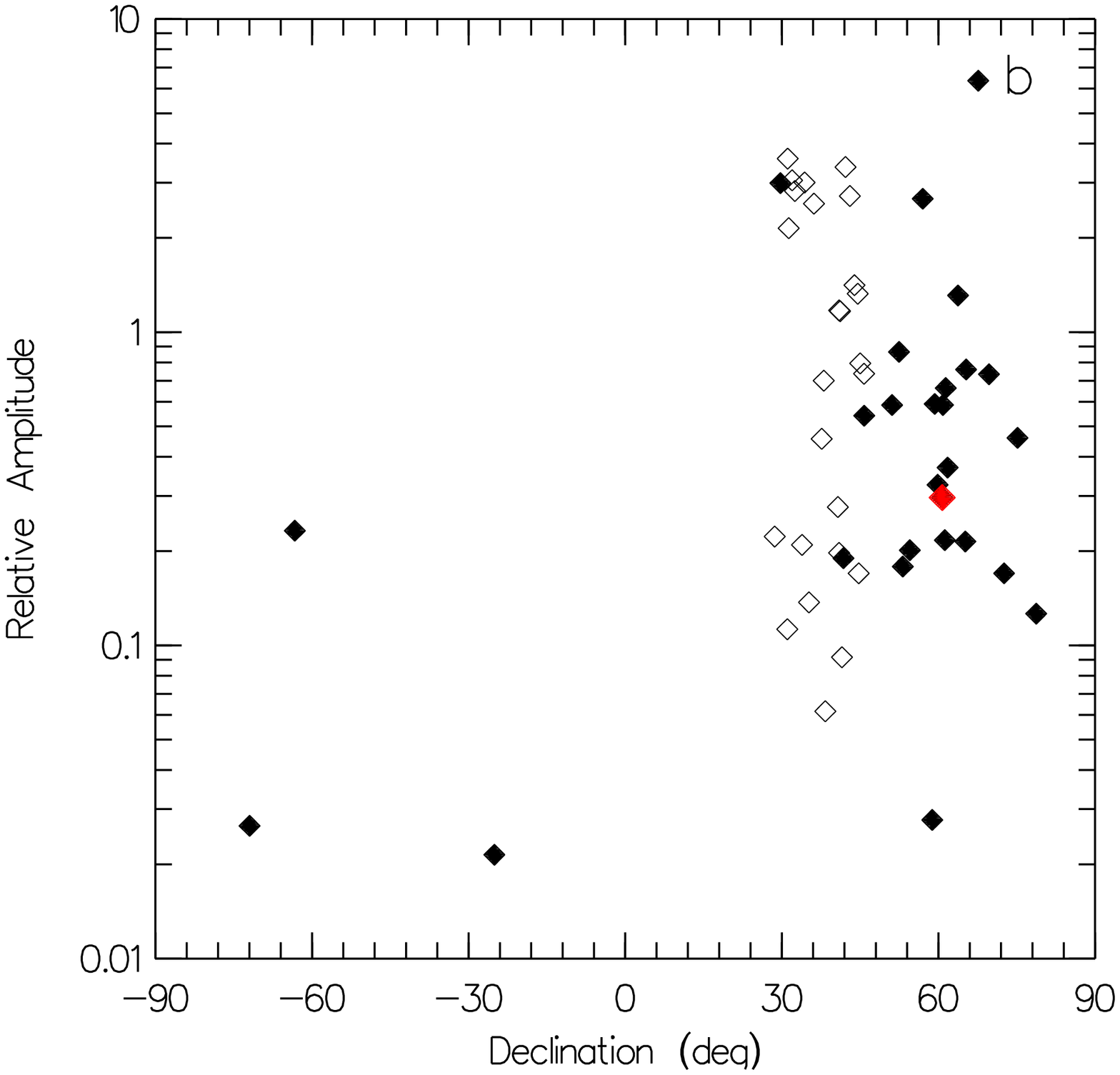} \\
\includegraphics[bb= 40 65 570 620,clip=true,height=\columnwidth,width=0.31\textheight,angle=-90]{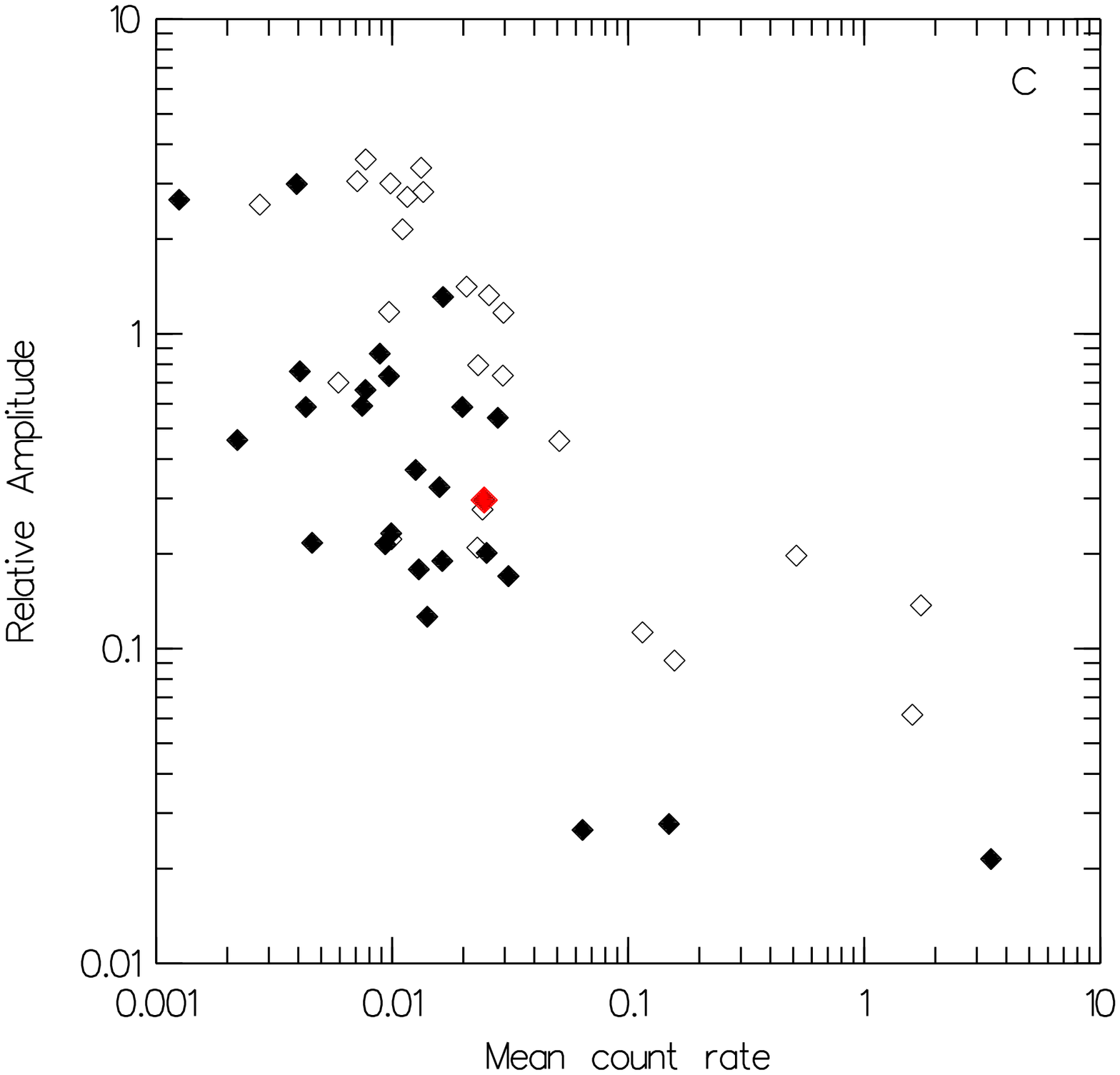} \\
\end{tabular}
\caption{Dependency of the relative amplitude of the 70\,d period modulation with Right Ascension (panel a), Declination (panel b) and mean count rate (panel c) for our sample of 55 MAXI sources. Count rates are in units of ph/s/cm$^{2}$. Filled symbols: light curves without gaps. Open symbols: light curves showing period gaps. The position of \gcas\ is marked by a large filled red lozenge.}
\label{precstudy}
\end{figure}
\end{center}

The power spectrum of the MAXI 2\,$-$20\,keV light curve of \gcas\ (Fig.~\ref{powerspecMAXI}) exhibits a relatively strong peak at P\,=\,70.5\,d, close to the $\approx$\,70\,d periodicity often present in the optical photometry \citep{smith2006}. Unfortunately, the best MAXI period is also identical to the 70\,d precession period of the {\it International Space Station} \citep{tomida2011} and this obviously casts doubt on the reality of this detection. We note in passing that no evidence of a 203.6\,d \gcas\ orbital period is present in the power spectrum of the MAXI data, thus confirming the spurious nature of the signal detected in the ASM data. In order to investigate to which extent the precession period of the ISS was imprinted in MAXI light curves, we extracted from the MAXI database light curves of the 43 X-ray sources located within 40\degree\ of \gcas, adding another 8 sources located within 60 degrees and finally 3 bright low-mass X-ray binaries at larger angular distances from \gcas\ (2S~0921-63, GX5-1 and LMCX-2). In all cases, we downloaded data averaged over one ISS orbital period and processed them applying the same MIDAS procedures as used for \gcas. In particular, we rebinned all individual orbit 2\,$-$20\,keV measurements in 5\,d bins and computed the LSPS of all light curves. We systematically discarded data obtained after MJD 56400 as those acquired at a later date seem affected by stronger errors. About half of the extracted light curves display periodic gaps that appear every 70\,d when the source moves away from the MAXI field of view. We found that most, if not all light curves are significantly modulated at the ISS precession period. Selecting the highest power in the period range of 66\,d to 78\,d, we computed the (half) amplitude of the best fitting sine and the mean count rate for all sources. Figs~\ref{precstudy} a \& b show that the relative (half) amplitude of the modulation does not seem to vary with the angular distance to \gcas. However, as shown on Fig.~\ref{precstudy} c, the relative amplitude steeply decreases with increasing source mean count rate. Such a behavior suggests that a background varying with the ISS precession period is at the origin of the spurious modulation. Light curves with gaps occurring every 70\,d display systematically higher relative amplitudes due to their peculiar periodic window function. \gcas\ remains continuously visible to MAXI at least until MJD 56400. The amplitude of the 70.5\,d modulation of \gcas\ is well within the range displayed by sources with similar X-ray brightness (see Fig.~\ref{precstudy} c). In addition, the $\approx$ 70\,d peak in the average LSPS of the 11 MAXI sources with fluxes, visibility and location on the sky similar to those of \gcas\ (Fig.~\ref{averagelsps}) is virtually the same as the 70.5\,d peak of \gcas\ itself (Fig.~\ref{powerspecMAXI}). We therefore conclude that the 70.5\,d modulation seen in the MAXI X-ray light curve of \gcas\ is mostly an artifact. We note, however, that close inspection of the optical and X-ray light curves indicates that X-ray power is maximum at times when strongest optical modulations occur so that a real correlation may exist blended with the underlying modulation produced by the precession of the ISS. This is supported by the strong close to zero time delay correlation between the optical and MAXI X-ray light curves observed early 2012. 

\end{document}